\documentstyle[12pt,psfig]{article}
\begin{document}

\thispagestyle{empty}

\vspace{4.0cm}
\begin{flushright} Preprint JINR\\
{\bf E2-97-133 }\\
hep-ph/9704333
\end{flushright}

\vspace{2.0cm}

\begin{center}
{\Large {\bf Analytic Model for the QCD Running Coupling }}

{\Large {\bf \vspace{0.3cm} }}

{\Large {\bf with Universal $\bar{\alpha}_s(0)$ Value } }\\

\vspace{2.cm}
{\large {\bf D.V. Shirkov and I.L. Solovtsov }}

\vspace{1.cm}
{\it Bogoliubov Laboratory of Theoretical Physics, JINR,\\
         Dubna, Moscow Region, 141980, Russia}
\end{center}

\vspace{2.cm}
\noindent
\begin{abstract}
We discuss the new model expression $\bar{\alpha}_{\rm an}(Q^2)$
recently obtained for the QCD running coupling with a regular
ghost-free behavior in the ``low $Q^2$" region. Being deduced from
the standard ``asymptotic-freedom" expression by imposing the
$Q^2$-ana\-ly\-ti\-city -- without any adjustable parameters -- it
obeys nice features:
(i)~The universal limiting value
$\bar{\alpha}_{\rm an}(0)=4\pi /\beta_0 \simeq 1.4$ expressed only
via group symmetry factors and independent of experimental estimates
on the running coupling $\bar{\alpha}_s(Q^2)$ (of QCD scale parameter
$\Lambda$). This value turns out to be stable with respect to higher
order corrections;
(ii)~Stability of IR behavior with respect to higher-loop effects;
(iii)~Coherence between the experimental $\bar{\alpha}_{\rm an}(M_{\tau}^2)$
value and integral information on IR  $\bar{\alpha}_s(Q^2)$ behavior
as extracted from jet physics data.
\end{abstract}

\newpage
\setcounter{page}{1}  The ghost-pole problem in the behavior of a running
coupling, being an obvious property of the geometrical progression, spoils a
physical discussion of the RG-summed perturbative QCD results in the
infrared (IR) region. To avoid it, one uses some artificial constructions
like the ``freezing of the coupling'' hypothesis.

Here, we are going to revive an old idea of combining the RG summation with
analyticity in the $Q^2$ variable. It was successfully used in the late 50's
for examining the QED ghost-pole issue~\cite{red,trio}. Quite recently, it
has been proposed for applying to the QCD case~\cite{jinr}.

The QED effective coupling $\bar{\alpha}(Q^2)$ being proportional to the
transverse dressed-photon propagator amplitude according to general
principles of local QFT satisfies the K\"all$\acute{{\rm e}}$n--Lehmann
spectral representation and, therefore, is an analytic function in the cut
complex $Q^2$ plane.

The ``analytization procedure'' elaborated in papers \cite{red,trio}
consists of three steps:

(I) To find an explicit expression for $\bar{\alpha}_{RG}(Q^2)$ in the
Euclidean region $Q^2>0$ by standard RG improvement of a perturbative input.

(II) To perform the straightforward analytical continuation of this
expression into the Minkowskian region ${\rm Re}\, Q^2<0$, ${\rm Im}\,
Q^2=-\epsilon$. To calculate its imaginary part and to define the spectral
density by $\rho_{RG} (\sigma ,\alpha)= {\rm Im}\bar{\alpha}_{RG}(-\sigma
-i\epsilon ,\alpha)$.

(III) Using the spectral representation [see Eq.~(\ref{spectral}) below]
with $\rho_{RG}$ in the integrand to define an ``analytically-improved''
running coupling $\bar{\alpha}_{{\rm an}}(Q^2)$ in the Euclidean region.

Being applied to $\bar{\alpha}(Q^2)$ in the one-loop ultra-violet (UV) QED
case, this procedure produced~\cite{trio} an explicit expression with the
following properties:

(a) it has no ghost pole,

(b) as a function of $\alpha$ at the point $\alpha=0$ it  possesses an
essential singularity

\noindent
$\sim~\exp(-3\pi/\alpha)$,

(c) in the vicinity of this singularity for real and positive $\alpha$ it
admits a power expansion that coincides with the  perturbation one (used as
an input),

(d) it has the finite UV limit $\bar{\alpha}(\infty,\alpha) = 3\pi $ that
{\it does not depend  on the experimental value} $\alpha\simeq 1/137$.

The same procedure being applied to the two-loop QED case yielded~\cite{trio}
a more complicated expression with the same essential features.

In the QCD case, to apply this technique to the strong running coupling, one
has to make two reservations.

First, as far as here $\bar{\alpha}_s(Q^2)$ is defined via a product of
propagators and a vertex function, there is a question about validity of the
spectral representation. Happily, this point has been discussed in paper~%
\cite{ilya}. As a result, one can use the K\"all$\acute{{\rm e}}$n-Lehmann
analyticity here, as well.

Second, in QCD, the running of coupling is, generally, connected with the
running of gauge. For simplicity, we assume that the $\overline{{\rm MS}}$
scheme is used (or the MOM scheme in the transverse gauge) when
$\bar{\alpha}_s$ is not influenced by the running of gauge.

To construct an analytic effective coupling in the QCD case, we start with
the leading-logs expression
\begin{equation}  \label{af}
\bar{a}^{(1)}(Q^2) = \frac{a}{1+a\beta_0\ln(Q^2/\mu^2)}= \frac{1}{
\beta_0\ln(Q^2/\Lambda^2)}
\end{equation}
with $\bar{a}(Q^2)=\bar{\alpha}_s(Q^2)/4\pi$ and $\beta_0=11-(2/3)n_f$, the
one-loop coefficient, and with the spectral representation
\begin{equation}  \label{spectral}
\bar{a}_{{\rm an}}(Q^2)= \frac{1}{\pi} \int_0^\infty d\sigma \frac{\rho
(\sigma ,a)}{\sigma+Q^2-{\rm i}\epsilon}\, .
\end{equation}
According to step (II) of the outlined procedure, we define the spectral
function in the one-loop approximation
\begin{equation}  \label{rho1}
\rho^{(1)}_{RG}(\sigma ,a)=\frac{a^2\beta_0\pi}{{(1+ a\beta_0 L)}^2+{
(a\beta_0\pi)^2}}\, ,\quad L=\ln\frac{\sigma}{\mu^2}.
\end{equation}
Note that the RG invariance of $\bar{a}_{{\rm an}}$ defined via Eq.(\ref
{spectral}) is provided by the scaling property of the spectral function
\begin{equation}  \label{rho-sc}
\rho (\sigma/\mu^2 ,a)=\tilde{\rho}(\ln \sigma/\Lambda^2),~~ \Lambda^2
=\mu^2\exp (-1/a \beta_0)~.
\end{equation}
Substituting $\rho^{(1)}_{RG}$ into Eq.~(\ref{spectral}) we get~\cite{jinr}
\begin{equation}  \label{a1}
\bar{a}^{(1)}_{{\rm an}}(Q^2)= \frac{1}{\beta_0}\left[\frac{1}{\ln
Q^2/\Lambda^2}\,+\,\frac{\Lambda^2}{\Lambda^2-Q^2}\right]~,
\end{equation}
where we used the QCD scale parameter defined as in Eq.(\ref{rho-sc}).
However, to identify $a$ with $a(\mu^2)$, the running coupling value at $%
Q^2=\mu^2$, we have to change this definition for $\Lambda^2=\mu^2\exp[%
-\phi(a(\mu^2)\beta_0)],$ where the function $\phi(z)$ satisfies the
equation $z=1/{\phi(z)}-1/[\exp (\phi (z))-1]$.

It is clear that the ``analytic'' coupling constant, Eq.~(\ref{a1}), has no
ghost pole at $Q^2=\Lambda^2$, and its IR limiting value
$\bar{\alpha}_s^{(1)}(0)= 4\pi/\beta_0 $ depends only on group factors.
Numerically, for
$n_f=3$, we have $\bar{\alpha}_{{\rm an}}^{(1)}(0)=4\pi/9\simeq 1.398$.

Usually, we are accustomed to the idea that theory supplies us with a set of
possible curves for $\bar{\alpha}_s(Q)$ and one has to fix the ``physical
one'' by comparing with experiment. Here, Eq.~(\ref{a1}) describes a family
of possible curves for $\bar{\alpha}_{{\rm an}}(Q^2)$ forming a bundle with
the same common limit at $Q^2=0$ as it is shown in Fig.~1.

\begin{figure}[tbp]
\centerline{\ \psfig{file=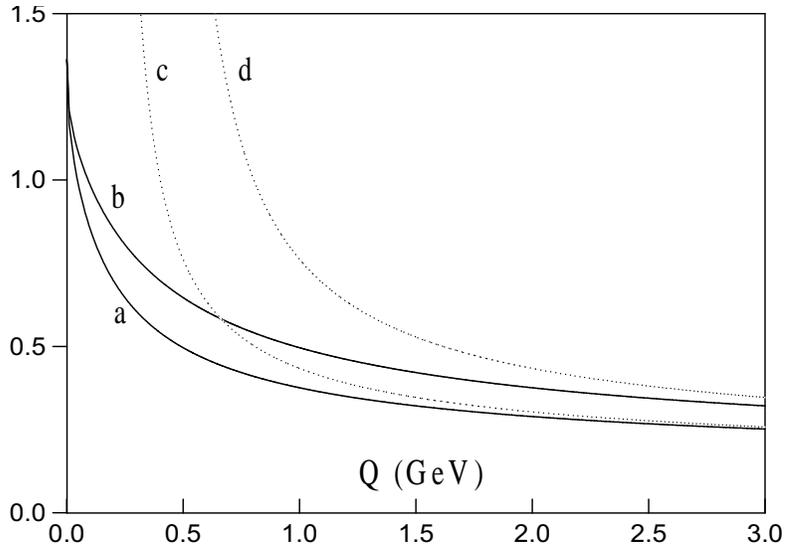,width=10.7cm}}
\caption{{\it The bundle of analytic solutions. The curves (a) and (b) are
the 1-loop analytic $\bar{\alpha}_s$ for $\Lambda=$200 MeV and 400 MeV,
respectively; the curves (c) and (d) show the corresponding perturbative
result.}}
\label{fig1}
\end{figure}

Another important virtue of Eq.~(\ref{a1}) is that the analytic behavior in
the IR domain is provided by a nonperturbative contribution $\simeq \exp
(-1/a\beta_0)$.

To analyze the two-loop case, let us start with $\bar{a}^{(2)}(Q^2)$ written
down in the form
\begin{equation}  \label{af2}
\bar{a}^{(2)}_{RG}(Q^2)=\frac{1}{\beta_0 l + \beta_0 b_1 \ln (1+l/b_1)}~ ,
\quad l=\ln\frac{Q^2}{\Lambda^2}\, ,
\end{equation}
where $b_1=\beta_1/(\beta_0)^2$ and $\beta_1=102-38/3n_f$ is the two-loop
coefficient. This expression corresponds to the result of exact integration
of the two-loop differential RG equation resolved by an iteration and
generates a popular two-loop formula with $\ln l/l^2$ term.

For the spectral density, we get
\begin{equation}  \label{rho2}
\beta_0 \rho^{(2)}_{RG} (L)\,=\,\frac{I(L)}{R^2(L)\,+\,I^2(L)}\, , \quad
L=\ln\frac{\sigma}{\Lambda^2}
\end{equation}
with
\begin{eqnarray}  \label{ri}
R(L)&=&L+b_1\ln \sqrt{\left(1+\frac{L}{b_1}\right)^2+ \left(\frac{\pi}{b_1}
\right)^2}~, \\
I(L)&=&\pi+b_1{\rm arccos}\frac{b_1+L} {\sqrt{\left(b_1+L\right)^2+\pi^2}}~.
\nonumber
\end{eqnarray}

Now, to obtain $\bar{a}^{(2)}_{{\rm an}}(Q^2)$, one has to substitute Eq.~(%
\ref{rho2}) into the r.h.s. of Eq.~(\ref{spectral}). However, the integral
expression thus obtained is too complicated for presenting in an explicit
form as the integration result differs from the used input $\bar{a}%
^{(2)}_{RG}$ not only by the pole term ``subtracting" the ghost pole (as in
the one-loop case), but also by an integral along the unphysical cut $%
~~0<Q^2<\Lambda^2\exp(-b_1)~~$ ``born" by the log-of-log dependence. For a
quantitative discussion we have to use numerical calculation.

Nevertheless, for a particular value at $Q^2=0$ we can make two important
statements. First, the IR limiting coupling value $\bar{\alpha}_{{\rm an}%
}(0)\,$, generally, does not depend on the scale parameter $\Lambda$. This
is a consequence of RG invariance (compare with Ref.\cite{gml}) and in our
case follows directly from Eq.(\ref{rho-sc}). Second, the IR limiting
coupling value is defined by the one-loop approximation, that is in the
two-loop case coincides with the one-loop case: $\bar{\alpha}_{{\rm an}%
}^{(2)}(0)\,=\,\bar{\alpha}_{{\rm an}}^{(1)}(0)\, \equiv \bar{\alpha}_{{\rm %
an}}(0)$.

To obtain a simple proof for the two-loop case, it is convenient to express
the difference $\bar{\alpha}_{{\rm an}}^{(1)}(0)-\bar{\alpha}_{{\rm an}%
}^{(2)}(0)$ via the imaginary part of the integral
\begin{equation}  \label{delta}
\Delta\,=\,\int_C\,\frac{dz}{z}\,\frac{\ln (1\,+\,z)} {z\,+\,\ln (1\,+\,z)}
\end{equation}
with the contour $C$ defined by $-\infty <{\rm Re} z < +\infty ,\, {\rm Im}
z=- \pi/ b_1 $. As far as the integrand is an analytic function in the
half-plane below the contour C, we conclude that $\Delta=0$. Note also that
the universality of $\bar{a}_{{\rm an}}(0)$ follows directly from the
procedure of constructing the analytic coupling. Indeed, our two-loop input
Eq.~(\ref{af2}) has the ghost pole at $Q^2=\Lambda^2$ and the unphysical cut
mentioned above. The analytization procedure removes these parasitic
singularities by two compensation terms. The term that removes the pole
gives the contribution $\Delta\bar{a}(0)_{{\rm pole}}=1/(2\beta_0)$. The
second term that compensates the cut can be expressed via the discontinuity
of function (\ref{af2}) on this cut and presented as
\begin{equation}  \label{cut-a}
\Delta\bar{a}(0)_{{\rm cut}}\,=\, \frac{1}{\beta_0}\,\int_0^\infty\,\frac{dz%
}{(z+1-\ln\, z)^2\,+\,\pi^2}\, ,
\end{equation}
which equals $1/(2\beta_0)$ and we again obtain the universal value of $\bar{%
a}_{{\rm an}}(0)=1/\beta_0$. Thus, in contrast to perturbation theory, where
the many-loop corrections change the IR behavior of the running coupling
significantly, our analytic coupling has a stable IR limit. The
analytization procedure removing all unphysical singularities leaves us with
the physical cut which is mainly described by the one-loop contribution.

Note also that the universality of the value of $\bar{a}_{{\rm an}}(0)$ is
not simply a matter of approximate resolution, Eq.~(\ref{af2}), of the exact
RG solution -- it has a deeper ground. This fact can be established in a
more general context, e.g., by considering the analytic properties, given by
the analytization prescription, in the complex $1/a$-plane. The details of
this reasoning are rather lengthy and will be published elsewhere.

Thus, the $\bar{\alpha}_{{\rm an}}(0)$ value, due to the RG invariance, is
independent of $\Lambda$ and, due to the analytic properties, independent of
higher corrections. This means that the causality (=analyticity) property
brings the feature of the {\sf universality}.

Here, we mean also that the whole shape of the $\bar{\alpha}_{{\rm an}}(Q^2)$
evolution turns out to be reasonably stable with respect to higher
corrections. The point is that the universality of $\bar{\alpha}_{{\rm an}%
}(0)$ practically gives rise to stability of the ${\bar{\alpha}_{{\rm an}%
}^{(\ell)}(Q^2)}$ behavior with respect to higher correction in the whole IR
region. On the other hand, this stability in the UV domain is a reflection
of the property of asymptotic freedom. As a result, our analytic model obeys
approximate ``higher loops stability" in the whole Euclidean region.
Numerical calculation (performed in the $\overline{{\rm MS}}$ scheme for
one-, two-, and three-loop cases with $n_f = 3$) reveals that ${\bar{\alpha}%
_{{\rm an}}^{(2)}(Q^2)}$ differs from ${\bar{\alpha}_{{\rm an}}^{(1)}(Q^2)}$
within the 10\% interval and ${\bar{\alpha}_{{\rm an}}^{(3)}(Q^2)}$ from ${%
\bar{\alpha}_{{\rm an}}^{(2)}(Q^2)}$ within the 1\% limit. This fact is
demonstrated in Fig.~2

\begin{figure}[tbp]
\centerline{\ \psfig{file=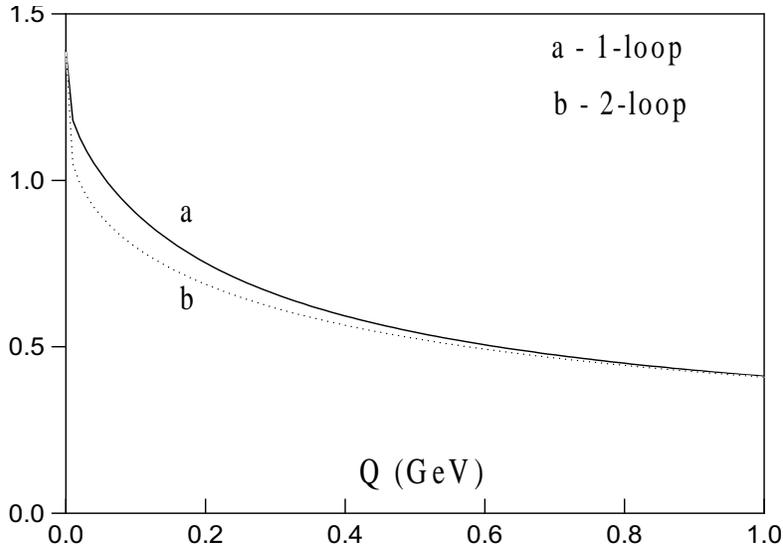,width=10.7cm}}
\caption{{\it ``Higher loop stability" of the analytic solution. The
normalization point is on the $\tau$ lepton scale: $\bar{\alpha}%
_s(M_{\tau}^2)=$0.34. Practically, the curve ${\bar{\alpha}_{{\rm an}%
}^{(3)}(Q^2)}$ coincides with ${\bar{\alpha}_{{\rm an}}^{(2)}(Q^2)}$ and we
do not plot it in the figure. }}
\label{fig2}
\end{figure}

It seems that the IR stability is an intrinsic feature of a non-analytic (in
$\alpha_s$) contribution. To illustrate this thesis, consider a recent IR
modification, ${\bar{\alpha}_B(Q^2)}$, for the QCD running coupling by
Badalian and Simonov~\cite{sim}.  These authors have studied a
non-perturbative contribution to the QCD running coupling on the basis of
the general background formalism using nonperturbative background
correlators as a dynamic input. They came to the conclusion that these
effects can practically be described by introducing an effective gluonic
mass $M_B=\sqrt{2\pi \sigma}$ defined by string tension $\sigma$ into all
``gluonic logarithms": $\ln Q^2 \to \ln (Q^2 +M_B^2) $. This means that ${%
\bar{\alpha}(Q^2)}$ essentially slows down its evolution (i.e., freezes)
around $Q \simeq M_B$. Their numerical estimate gave $M_B \simeq 1.5~{\rm GeV%
}$. In practice, this yields the difference between one-- and two--loop
results in the IR region of $15\%$ order.

In our calculations, we used $n_f=3$ as an average quark number. This seems
to be reasonable in the low energy region $Q< 2 \,{\rm GeV} \simeq m_c$. For
a more realistic description of the $\bar{\alpha}_s(Q)$ evolution in the
whole Euclidean domain, one should take into account quark thresholds. To
this end, one usually applies a matching procedure, changing abruptly the
number of active quarks at an ``effective" threshold $Q=\xi m_q$ with some
matching parameter $\xi \simeq 1$. Evidently, any procedure of that sort
violates the $Q^2$ analytic properties. On the other hand, these properties
could be preserved by the ``smooth matching" algorithm devised~\cite{sh92}
on the base of an explicitly mass-dependent RG formalism ascending to
Bogoliubov.  This algorithm has been recently used~\cite{mikh} for the
precise analysis of the $\bar{\alpha}_s(Q)$ evolution in the $3 \,{\rm GeV}
< Q < 100\, {\rm GeV}$ interval.

In the present context it could be applied also in the $Q \leq \Lambda$
domain. However, this would change the $\bar{\alpha}_{{\rm an}}(Q^2)$
behavior in the ``very low $Q$" region only slightly as far as the limiting $%
\bar{\alpha}_{{\rm an}}(0)$ value depends on $\beta_0(n_f^*)$ with the
effective quark number $n_f^*$ related to the ghost-pole position $\Lambda$.
As it is generally accepted on the base of DIS data, in the $\overline{{\rm %
MS}}$ scheme $\Lambda_{n_f=3}(2-{\rm loop}) \simeq 250\div 350\, {\rm MeV}$
which is quite above of the strange quark mass. This means that the use of
value $n_f^*=3$ is justified.

Analytic properties of the running coupling are important from the point of
view of phynomenological applications, for example, for the description of
the inclusive decay of the $\tau $ lepton. To this end, one usually
transforms the initial expression for the $R_\tau $ ratio to the integral
form in the complex $q^2$-plane (see, e.g.,~\cite{braaten-nar-pich}). This
transformation based on analytic properties mentioned above which are
violated in the standard perturbative consideration and maintained within
our method. Note also that referring to ``low-$Q$'' data, like those of $%
\tau $-lepton decay, one should distinguish between QCD scale $\Lambda ^{%
\overline{{\rm MS}}}$ in the usual RG solution taken in the $\overline{{\rm %
MS}}$ scheme and $\Lambda ^{{\rm an}}$ corresponding to our analytic
expression. For instance, in the one-loop case, to the $\bar \alpha
_s(M_\tau ^2)=0.34$ there correspond $\Lambda _3^{{\rm an}}=280\,{\rm MeV}$
as compared with $\Lambda _3^{\overline{{\rm MS}}}=230\,{\rm MeV}$.

The idea that the QCD running coupling can be frozen or finite at small
momenta has been considered in many papers (see, e.g., the discussion in~%
\cite{M-Stev}). There seems to be experimental evidence in favor of this
behavior of the QCD coupling in the IR region. As an appropriate object for
comparison with our construction, we use the average
\begin{equation}  \label{Eq.aq}
A(Q)\,=\,\frac{1}{Q}\,\int_0^Q\,dk\,{\bar{\alpha}_s}(k^2)
\end{equation}
\noindent
that people manage to extract from jet physics data. Empirically, it has
been claimed that this integral at $Q\simeq 1\div 2$ GeV turns out to be a
fit-invariant quantity. For it there is an estimate: $A(2\,\,{\rm GeV})=0.52
\pm 0.10$~\cite{Dok-Webb}. Our results for $A(2\,\,{\rm GeV})$ obtained by
the substitution $\bar{\alpha}^{(1)}_{{\rm an}}$ and $\bar{\alpha}^{(2)}_{%
{\rm an}}$ into Eq.~(\ref{Eq.aq}) for some values of the running coupling at
the normalization point $\mu^2=M_{\tau}^2 $ are summarized in the Table.

\begin{center}
\begin{tabular}{|c|c|c|c|}
\hline
$\bar{\alpha}_{{\rm an}}(M_{\tau}^2)$ & 0.34 & 0.36 & 0.38 \\ \hline
~~~$A_{{\rm 1-loop}}(2\,{\rm GeV})$~~~ & ~~~0.50~~~ & ~~~0.52~~~ & ~~~0.55~~~
\\ \hline
~~~$A_{{\rm 2-loop}}(2\,{\rm GeV})$~~~ & ~~~0.48~~~ & ~~~0.50~~~ & ~~~0.52~~~
\\ \hline
\end{tabular}
\end{center}

Note here that a nonperturbative contribution, like the second term on the
l.h.s. of Eq.~(\ref{a1}), reveals itself even at moderate $Q$ values by
``slowing down" the rate of $\bar{\alpha}_s(Q)$ evolution. For instance, in
the vicinity of the $c$ and $b$ quark thresholds at $Q=3\,{\rm GeV}$ it
contributes about 4\%, which could be essential for the resolution of the
``discrepancy" between ``low-$Q$" data and direct $Z_0$ measurement for $%
\bar{\alpha}_s(M_Z)$.

\vspace{0.2cm} In this letter, we have argued that a possible way to resolve
the ghost-pole problem for the QCD running coupling can be found by imposing
the K\"all$\acute{{\rm e}}$n-Lehmann $Q^2$-analyticity which reflects the
causality principle of QFT. The analytic behavior in the IR region is
restored by a nonperturbative contribution. The procedure of constructing
the analytic running coupling is not unambiguous~\cite{kff}. We have
considered the simplest way which does not require any additional parameters
and operates only with $\Lambda$ or a value of the coupling at a certain
normalization point. The requirement of analyticity yields significant
modification in the IR and intermediate domains and leads to the universal
value of $\bar{\alpha}_{{\rm an}}(0)$. In this paper we have obtained the
analytically-improved result $\bar{\alpha}_{{\rm an}}(Q^2)$ for the QCD
running coupling that turns out to be quite stable with respect to
higher-order corrections for the whole interval of $Q^2$ and agrees with low
energy experimental evidence for the IR-finite behavior.

Our construction does not contain adjustable parameters. This is due, in
particular, to the convergence of a nonsubtracted spectral integral, that
is, with the asymptotic freedom property. Here, analyticity plays the role
of a bridge between regions of small and large momenta. The idea that
``analyticity is the key factor relating high energy to low energy'' has
recently been emphasized by Nishijima~\cite{nish} in the context of
connection between the asymptotic freedom and color confinement. However, in
our approach this connection is not so direct. If, e.g., we admit (see Ref.~%
\cite{erhard}) the possibility of a UV fixed point for the QCD effective
coupling at some small value of $\alpha_s\leq 0.05$, then we arrive at a
modification of the UV behavior with a power instead of $(\ln \sigma)^{-2}$
decrease of the spectral function. So, in that case we can also use a
nonsubtracted spectral representation.

As far as it is difficult to present an explicit analytic expression for ${%
\bar{\alpha}_{{\rm an}}^{(2)}(Q^2)}$, for the need of QCD practitioners, we
propose an approximate formula. It can be obtained by the method of
subtraction of unphysical singularities if one takes into account the
explicit expression for the term that removes the ghost pole and an
approximate expression in the form of the first term of the power expansion
for the term that removes the unphysical cut. The corresponding expression
reads
\begin{equation}  \label{approximation1}
\beta_0\tilde{a}^{(2)}(Q^2)\simeq \beta_0\bar{a}^{(2)}_{RG}(Q^2) +\frac{1}{2}%
\;\frac{1}{1-Q^2/\Lambda^2}-\frac{\Lambda^2}{Q^2}C_1\, ,
\end{equation}
where $\bar{a}^{(2)}_{RG}(Q^2)$ is defined by (\ref{af2}) and, for $n_f=3$, $%
C_1=0.035$. Expression~(\ref{approximation1}) approximates the two-loop
analytic coupling with an accuracy less than 0.5\% in the region $%
2.5\,\Lambda <Q<3.5\,\Lambda$, practically coincides with the exact formula
for larger values of momenta, and, therefore, can be used in the analysis of
many experimental data. If there is necessary to consider a very small
momentum, like Eq.~(\ref{Eq.aq}), we suggest another approximate formula
which can be written in the form of Eq.~(\ref{a1}) with substitution,
instead $Q^2/\Lambda^2$, the expression $\exp[\ln\,{Q^2}/{\Lambda^2}+b_1 \ln%
\sqrt{\ln^2\,{Q^2}/{\Lambda^2}+4\pi^2}]. $ For $Q\geq \Lambda$, the accuracy
of this approximation is less than 5\% and yields only 3\% error into the $%
A(2\,{\rm GeV})$ value.

\vspace{0.2cm} The authors would like to thank A.M.~Baldin, H.F.~Jones,
A.L.~Kataev, K.A.~Milton, V.A.~Rubakov and O.P.~Solovtsova for interest in
the work and useful comments. Partial support of D.Sh. by INTAS 93-1180 and
RFBR 96-15-96030 grants and of I.S. by the US NSF grant PHY-9600421 is
gratefully acknowledged.


\begin{thebibliography}{99}
\bibitem{red}  P. Redmond, Phys. Rev. {\bf 112} 1404 (1958);  P. Redmond and
J.L. Uretsky, Phys. Rev. Lett. {\bf 1} 147 (1958).

\bibitem{trio}  N.N. Bogoliubov, A.A. Logunov and D.V. Shirkov,  Sov. Phys.
JETP {\bf 37}(10) 574 (1960).

\bibitem{jinr}  D.V.~Shirkov and I.L.~Solovtsov, JINR Rapid Comm.,  No.
2[76]-96, 5, hep-ph/9604363.

\bibitem{ilya}  I.F. Ginzburg and D.V. Shirkov, 
Sov. Phys. JETP {\bf 22} 234 (1966).

\bibitem{gml}  M.~Gell-Mann and F.~Low, Phys.Rev., {\bf 95} 1300 (1954).

\bibitem{sim}  Yu.A. Simonov, Yad. Fizika, {\bf 58} 1139  (1995); A.M.
Badalian and Yu.A. Simonov, loc. cit. (to be  published).

\bibitem{sh92}  D.V.~Shirkov, Nucl. Phys. B {\bf 371} 476 (1992).

\bibitem{mikh}  D.V. Shirkov and S.V. Mikhailov, Z. Phys. C {\bf 63} 463
(1994).

\bibitem{braaten-nar-pich}  E.~Braaten, S.~Narison and A.~Pich, Nucl. Phys.
B {\bf 373} 581 (1992).

\bibitem{M-Stev}  A.C. Mattingly and P.M. Stevenson, Phys. Rev. D {\bf 49}
437 (1994).

\bibitem{Dok-Webb}  Yu.L.~Dokshitzer and B.R.~Webber, Phys. Lett. B {\bf 352}
451 (1995); Yu.L.~Dokshitzer, V.A.~Khoze and S.I.~Troyan, Phys. Rev. D {\bf %
53} 89 (1996).

\bibitem{kff}  D.A.~Kirzhnits, V.Ya.~Fainberg and E.S.~Fradkin, Sov. Phys.
JETP {\bf 11} 174 (1960).

\bibitem{nish}  K.~Nishijima, Czech. J. Phys. {\bf 46} 1 (1996).

\bibitem{erhard}  A. Patrascioiu and E.~Seiler,
Phys. Rev. Lett. {\bf 74} 1920 (1995); {\em ibid} {\bf 74} 1924 (1995).
\end{thebibliography}
\end{document}